\documentclass[amsmath,amssymb,aps,prb,twocolumn,superscriptaddress]{revtex4-2}
\usepackage[utf8]{inputenc}
\usepackage[normalem]{ulem}
\usepackage{physics}
\usepackage{epsfig,subfigure,amsmath,float,xcolor}
\newcommand{\sig}{\sigma}
\newcommand\beq{\begin{equation}}
\newcommand\eeq{\end{equation}}
\newcommand\bes{\begin{subequations}}
\newcommand\ees{\end{subequations}}
\newcommand\bea{\begin{eqnarray}}
\newcommand\eea{\end{eqnarray}}
\newcommand\non{\nonumber}

\newcommand\ig{\includegraphics}

\newcommand\al{\alpha}

\newcommand\lam{\lambda}

\newcommand\sfig{\subfigure}

\newcommand\vk{{\bf k}}

\newcommand{\RNum}[1]{\uppercase\expandafter{\romannumeral #1\relax}}

\begin{document}
\title{Josephson junctions of topological nodal superconductors}
\author{Ranjani Seshadri}
\email{ranjanis@post.bgu.ac.il}
\affiliation{Department of Physics, Ben-Gurion University of the Negev,
Beer-Sheva 84105, Israel}
\author{Maxim Khodas}
\email{maxim.khodas@mail.huji.ac.il}
\affiliation{Racah Institute of Physics, Hebrew University of Jerusalem,
Jerusalem 91904, Israel}
\author{Dganit Meidan}
\email{dganit@bgu.ac.il}
\affiliation{Department of Physics, Ben-Gurion University of the Negev,
Beer-Sheva 84105, Israel}
\date{\today}
\begin{abstract}
Transition metal dichalcogenides (TMDs) offer a unique platform
to study unconventional superconductivity, owing to the presence of strong 
spin-orbit coupling and a remarkable stability  to an in-plane magnetic field. 
A recent study found that when an in-plane field applied to a superconducting
monolayer TMD is increased beyond the Pauli critical limit, a quantum phase transition
occurs into a topological nodal superconducting phase which hosts Majorana flat bands.
We study the current-phase relation of this nodal superconductor in a Josephson
junction geometry. We find that the nodal superconductivity is associated with an
energy-phase relation that depends on the momentum transverse to the current direction,
with a $4\pi$ periodicity in between pairs of nodal points. We interpret this
response as a result of a series of quantum phase transitions, driven by the
transverse momentum, which separate a topological trivial phase and two distinct
topologically non-trivial phases characterized by different winding invariants.
This analysis sheds light on the stability of the Majorana flat bands to
symmetry-breaking perturbations. 
\end{abstract}
 \maketitle
 
 %input{Intro}
\section{Introduction}
Recent advances in  fabrications techniques have made it possible to engineer high-quality
ultra-thin, multi-layer systems based on transition metal dichalcogenides (TMDs), with
individual layers held together by weak Van der Waals forces \cite{Wang2012,Geim2013}.
Some of these few-layered systems remain superconducting down to the monolayer limit 
\cite{Lu2015,Ugeda2015,Saito2016,Xi2016,Costanzo2016,Dvir2017,DelaBarrera2018,Sohn2018}.
%greatly stimulating the research of two-dimensional superconductivity. 
Such TMD-based systems have been proposed as a platform for controlled studies of intrinsic
or externally-induced unconventional superconductivity
\cite{Yuan2014,Zhou2016,Hsu2017,He2018,Fischer2018,Hsu2020,Cho2020,Kanasugi2020,
Glodzik2020,Hamill2021,Shaffer2020,Nayak2021,wan2021,ganguli2020,Wickramaratne2020,amit2020}.

Unlike bulk systems, many TMD monolayers such as 1H-NbSe$_2$ lack an inversion center.
This causes a spin-orbit splitting of electron bands, which polarizes the spin in the
out-of-plane direction. 
% Due to the basal mirror plane symmetry
% $\sigma_h$, the electron spins are polarized out-of-plane. 
The superconducting properties of such systems, termed Ising superconductors (SCs),
\cite{Yuan2014,Lu2015,Saito2016,Xi2016} are determined  by the spin-orbit coupling (SOC),
$\Delta_{\mathrm{SO}}$ typically exceeding the superconducting gap by few orders of magnitude.
In particular, Ising superconductivity is remarkably stable to the in-plane magnetic field,
$\mathbf{B} \perp \hat{z}$. The in-plain critical field, $B_c$ greatly exceeds the Pauli limit
\cite{Lu2015,Saito2016,Xi2016,Dvir2017,DelaBarrera2018,Sohn2018,Liu2018} and at zero
temperature is limited by the disorder  \cite{Bulaevskii976,Sosenko2017,Ilic2017,Mockli2020}.
% These properties together with their high tunnability, make TMD based-systems a
%compelling platform for controlled studies of  intrinsic or
% externally induced unconventional superconductivity \cite{Hamill2021,Cho2020}. 

While its presence explicitly breaks time-reversal (TR) symmetry $ \Theta $, the stability of 
the superconducting state to the in-plane field can be understood as resulting from a modified 
TR symmetry $T$ which is a combination of the TR symmetry $\Theta$ and basal-plane mirror symmetry
$M_z$  and is given by $T=M_z \Theta \tau_z $, where $\tau_{x,y,z}$
are the Pauli matrices in particle-hole Nambu space. This modified TR symmetry
protects the superconducting state \cite{Fischer2018} and gives rise to 
% , which commute with the Zeeman coupling to the  in-plane  field
% \cite{Fischer2018}. This symmetry gives rise to the 
field-induced triplet correlations \cite{Mockli2018}.

Recently it was predicted  that as the applied in-plane magnetic field exceeds the superconducting
gap, a monolayer Ising SC transitions into a nodal topological SC \cite{He2018}.
The formation of nodal points 
% along $\Gamma M$ high symmetry lines 
is protected by the  effective chiral symmetry for the particles moving
%MK along the direction 
perpendicular to the $\Gamma M$  line, which results from a combination of the modified
TR symmetry $T$ and particle-hole symmetry.
%MK changes
The nodal phase is  expected to be accompanied by 
%MK the formation of
Majorana flat bands along the armchair edges \cite{Zhao2013,Matsuura2013}, experimental indications
of which have been reported in \cite{Galvis2014,Nayak2021}. 

In this work we study the Josephson response of the nodal SC phase. We find that the nodal SC phase
is associated with an energy-phase relation dependent on the momentum transverse to the 
%MK junction, 
current direction with a distinct $4\pi$-periodic Josephson current for the transverse momenta
in-between the nodal points.
%in between the nodal points. 
We interpret this response as a  consequence of a series of topological transitions
resulting from the continuous change of the transverse momentum considered as the control parameter.
%MK driven by the transverse momenta, 
The nodal momenta define the boundaries %MK moment phases are 
between a trivial phase and two topologically distinct non-trivial phases with different
winding numbers. We further discuss the implications of these results on the stability of the
Majorana flat bands in the presence of a symmetry-breaking perturbation. 

The plan of this paper is as follows. We begin in Sec. \ref{sec:BdG} with a discussion of the
symmetries that dictate the form of the low-energy Bogoliubov-deGennes (BdG) Hamiltonian of an
Ising SC. In Sec. \ref{sec:top} we study the topological properties of the nodal SC phase and
calculate the corresponding invariants. The current-phase relation in a Josephson
junction made of two such nodal SCs is analyzed in Sec. \ref{sec:JJ} and the results are
presented in Sec. \ref{sec:res}, followed by a discussion of the underlying physical picture
in Sec. \ref{sec:disc}. We accompany these qualitative arguments with a detailed description
of edge states of an effective one-dimensional theory in appendix \ref{sec:edgemodes}, while the
effect of the magnetic field on the transition is  analyzed in appendix \ref{app:phase}.
The dependence of the Josephson current response on the junction barrier strength is briefly
discussed in appendix \ref{app:barr}.

\section{BdG Hamiltonian for a Nodal superconductor} \label{sec:BdG}
The band structure of TMD monolayers is constrained by the underlying crystalline symmetry with
the point group $D_{3h}$. The symmetry operations include basal mirror reflection, $M_z$ which
does not change the in-plane momentum and acts solely on the spin, $M_z = -i\sigma_z $.
%MK consist of out-of-plane mirror symmetry $M_z = -i\sigma_z $, a three-fold
In addition, the rotation $C_3= \left\{e^{-i\frac{\pi}{3}\sigma_z}|\vk\rightarrow\hat{R}_z(2\pi/3)\vk\right\}$
around the $z-$axis acts both on spin and the momentum, with
$\hat{R}_z(2\pi/3)$ being the spatial rotation by $2\pi/3 $ around the $z-$axis.
Finally, a vertical mirror passing via the high symmetry $\Gamma M$ line,
taken here to lie along the $y-$axis, $M_{x}=\left\{-i\sigma_x|k_x,k_y \rightarrow -k_x,k_y\right\}$.
 %MK Here $\hat{R}_z(2\pi/3)$ is the spatial rotation by $2\pi/3 $ around the $z$ axis. 

Inversion is not included in $D_{3h}$ group causing a finite Ising SOC.
Thanks to the $M_z$ symmetry, the electron spins are polarized out-of-plane.
%MK  Apart from these crystalline symmetries, many  TMD monolayers lack  an inversion center which 
%causes a spin orbit splitting of electron bands, which polarizes the spin in the out of plane direction.  
The vertical mirror symmetry operation, $M_x$ makes the Ising SOC odd under the momentum
reflection $k_x\rightarrow -k_x$. Correspondingly, the Ising SOC is even under $k_y\rightarrow -k_y$.

For definiteness, we consider the band structure of NbSe$_2$ monolayer with one band crossing the Fermi level.
Different crossings give rise to the hole pockets centered at the $\Gamma$, $K$ and $K'$ points. 
%MK \cite{He2018}. 
The strong SOC near the $K(K')$ points protects the superconductivity in these pockets which is
nearly unaffected by an in-plane magnetic field. Therefore, the analysis presented here relies
solely on the presence of the $\Gamma$-point centered pocket(s) and applies equally to other
systems such as gated MoS$_2$ \cite{Lu2015,Saito2016,Costanzo2016}, gated WS$_2$ \cite{Lu2018},
and metallic TaS$_2$ \cite{DelaBarrera2018}.
To understand the interplay between the SOC, in-plane field and superconductivity, it is
sufficient to focus on the low-energy Hamiltonian as these energy scales are smaller than
the typical Fermi energy. 

To describe the spectrum of the $\Gamma$-pocket we expand the low energy single-band Hamiltonian, 
\beq
H_0 = \xi(\vk)  \sig^0 + \lam(\vk) \sig^z .
\eeq
up to the third order in $k$. %reads, 
%MK $D_{3h}$ point-group symmetry of the underlying lattice described 
%above constrains the form of the momentum-space Hamiltonian to order $ k^3$, which takes the following form:
Here $\xi(\vk) = (k_x^2+k_y^2)/2m -\mu$, where % \label{eq:xi},
$\mu$ and $m$ are the chemical potential and the mass, respectively.
The Ising SOC term is given by 
\beq
\lam(\vk)  = \lam_I(k_x^3-3 k_x k_y^2), \label{eq:lamk}
\eeq 
where $\lam_I$ is the strength of the Ising SOC. As is clear from
Eq.\eqref{eq:lamk}, this form of SOC vanishes along the $\Gamma M$ lines  $k_x=0, \pm \sqrt{3}k_y$.
% Moreover, Ising SOC has the special property that it locks the spins of electrons with
% opposite momenta in the opposite out-of-plane directions. \cite{He2018}
Note that in the following $t a^2 =m^{-1} $ is set to one, where $ t$ is the tight binding hopping amplitude and $a $ is the lattice constant. 

In order to study the interplay between an in-plane magnetic field $B$ characterized by the
Zeeman energy $h = g\mu_B B/2$ applied for definiteness along the $x-$direction and an
$s$-wave spin-singlet superconducting pairing characterized by the gap, $\Delta$,
we consider the BdG Hamiltonian,

\beq
\mathcal{H}(\vk) = \xi(\vk) \tau^z   + \lam(\vk)  \sig^z+ h \tau^z  
\sig^x + \Re(\Delta) \tau^y \sig^y+ \Im(\Delta) \tau^x \sig^y. \label{eq:Hkk}
\eeq

The magnetic field explicitly breaks TR symmetry 
$\Theta =\left \{i\sigma_y K|\vk \rightarrow -\vk\right\} $,
where $K$ stands for complex conjugation. However, the combination of $\Theta$ and $M_z$
results in a modified TR symmetry,
$T=\Theta M_z \tau_z=\left\{ \sigma_x \tau_z K| \vk \rightarrow -\vk\right\} $,
which squares to $T^2=1 $.  Moreover, thanks to the $M_x$ mirror symmetry,
the Hamiltonian is even in $k_y $.  We, therefore have the symmetry
$T\mathcal{H}(\vk) T^{-1} =\mathcal{H}(-\vk) = \mathcal{H}(-k_x,k_y)$ for the motion
along $x-$direction when $k_y$ parameter is fixed, as is elaborated below.

\begin{figure}[htb]
\centering
\ig[height=7cm]{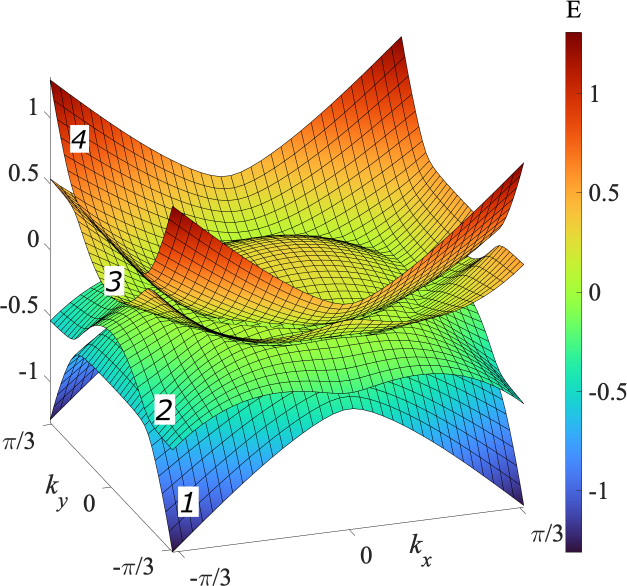}
\caption{The four band dispersion relation $E$ vs $\vk$ of the
Bogoliubov quasi-particles from the BdG Hamiltonian in Eq.\eqref{eq:Hkk}.
In this figure we have chosen $m=1$, $\mu=0.25$, $\lam_I=0.15$, $h=0.06$
and $\Delta=0.02$. The four bands are labelled $1-4$ in increasing order of 
energy. The surface plot of band $3$ is shown in Fig. \ref{fig:surf_H6Del2}.}
\label{fig:4bands}
\end{figure}

\begin{figure}[htb]
\centering
\begin{center}
{\ig[height=7cm]{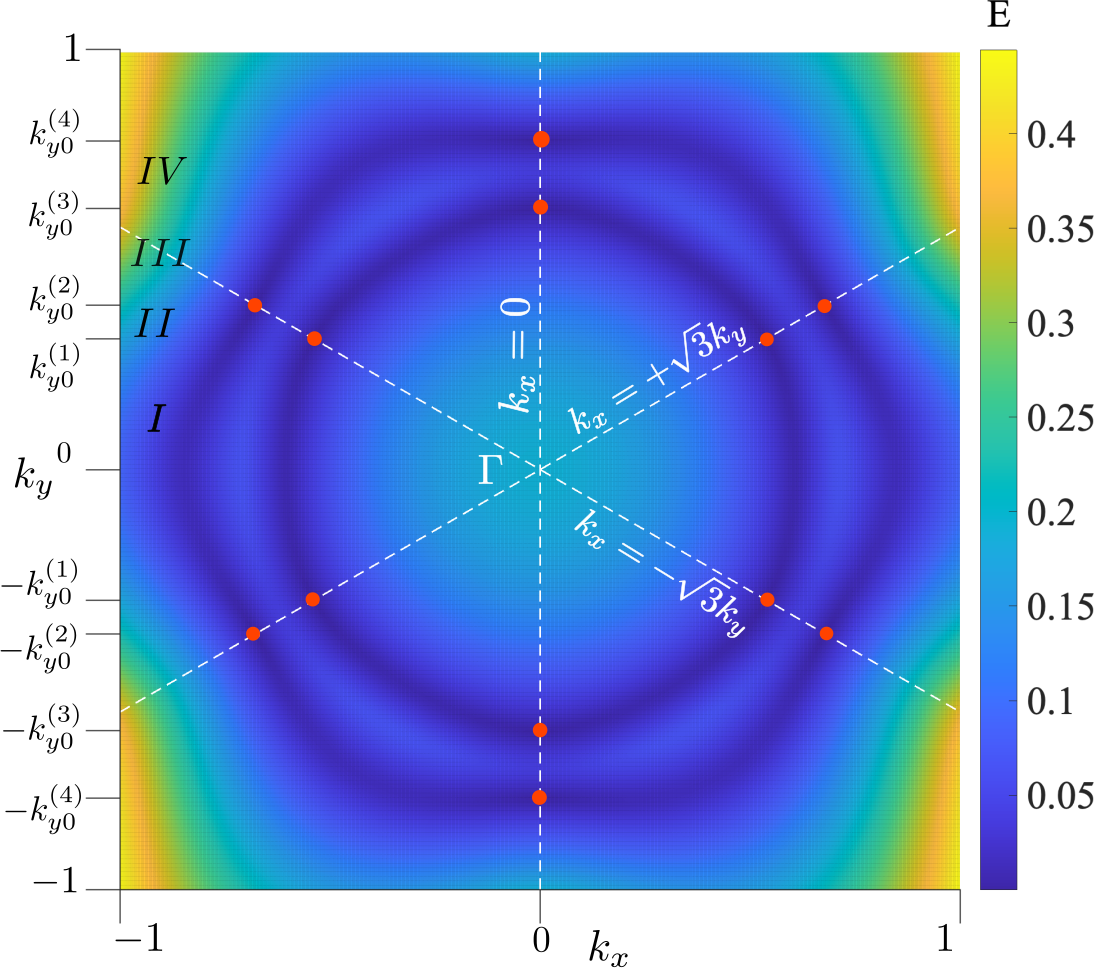}}
\end{center}
\caption[Band 3]{Color plot of the lower positive energy band obtained (labelled $3$ in
Fig. \ref{fig:4bands}). The spectral gap closes at twelve nodal points in the
Brillouin Zone. These are shown as red dots and lie on the  $\lam(\vk) = 0$ lines
(white, dashed lines). The $y-$momenta $\pm k_{y0}^{(j)}$ mark the boundaries between
the topological and non-topological phases which we cross as we change $k_y$ continuously.}
\label{fig:surf_H6Del2}
\end{figure}

The dispersion relation of the Bogoliubov quasi-particles inferred from Eq.~\eqref{eq:Hkk} 
forms the four bands as shown in the Fig. \ref{fig:4bands} for selected values
of the parameters ($m=1$, $\mu=0.25$, $\lam_I=0.15$, $h=0.06$ and $\Delta=0.02$). 
Figure \ref{fig:surf_H6Del2} shows the lower positive-energy band (labelled $3$ in
Fig. \ref{fig:4bands}) as a color plot.

Due to the modified TR symmetry, superconductivity survives \cite{Fischer2018}
when the magnetic field exceeds far beyond the Pauli limit.  At $h=|\Delta|$, a
quantum phase transition occurs, accompanied by the closing of the band gap at six
discrete nodal points as detailed in the App.~\ref{app:phase}.  At yet stronger
magnetic field exceeding the superconducting gap, $h>|\Delta|$ each nodal point splits
in two, as shown by the red dots in Fig.\ref{fig:surf_H6Del2}. The resulting twelve
nodal points lie along high-symmetry $\Gamma M$ lines, marked in white.

We show below that this nodal SC phase, which arises from the model's non-trivial topological
properties \cite{He2018}, is accompanied by a distinctive signature in the Josephson
current response, which can be used as a probe to experimentally detect the topological phase. 

\section{Symmetries and Topological properties}\label{sec:top}
% 		\begin{figure}\label{fig:dispersion}
% 	%	\includegraphics[width=.5\textwidth]{data/dispersion_delta_002_h_006_lambda_015_mu_025_D_0885.png}
% 	\end{figure}
The  origin of the nodal topological SC phase can be understood by
considering the family of one-dimensional (1D) Hamiltonians obtained by setting
the momentum  $k_y $  as a parameter in the BdG Hamiltonian of Eq.~\eqref{eq:Hkk}.
We denote this Hamiltonian as $\mathcal{H}_{k_y}^{1D }(k_x)$. While the magnetic
field explicitly breaks TR symmetry, the family of 1D Hamiltonians has an
emergent TR symmetry
%MK \Theta has been introduced in Intro 
%given by 
%$\Theta = \sigma_x\tau_z K $
$\Theta$ such that $ \Theta \mathcal{H}_{k_y}^{1D } (k_x) \Theta^{-1} =
\mathcal{H}_{k_y}^{1D } (-k_x) $. 
In addition to the particle-hole symmetry $\Xi = \tau_x K$, the family of 1D Hamiltonians,
$\mathcal{H}_{k_y}^{1D }(k_x)$ have a chiral symmetry $\mathcal{C}= \sigma_x \tau_y$
and fall under symmetry class BDI. 
As the parameter $k_y$ is varied, the gap closes and reopens at the spectral nodes. 
This closing and reopening of the gap is accompanied by a transition from a topologically
trivial to a topologically non-trivial phase. %MK, as explained below. 
To study the topological properties of this family of one-dimensional
Hamiltonians, we rotate to the chiral basis $\tilde{\mathcal{H}}^{1D}_{k_y} = U \mathcal{H}_{k_y}^{1D } U^{-1}$ with $U=e^{-i\pi/4 \tau_y}e^{-i\pi/4 \sigma_x\tau_z} $.
The rotated Hamiltonian
$\tilde{\mathcal{H}}^{1D}_{k_y}$ can be written as
\begin{eqnarray}
\tilde{\mathcal{H}}^{1D}_{k_y} (k_x) = \left(\begin{array}{cc}
0&\mathcal{Q}_{k_y}(k_x)\\
\mathcal{Q}_{k_y}^\dag(k_x)&0
\end{array}\right) 
\end{eqnarray}
where:
\begin{eqnarray}
\mathcal{Q}_{k_y}(k_x) &=& \left(\begin{array}{cc}
\xi_{k_y} (k_x)&h+	i\lambda_{k_y}(k_x) -\Delta 
\\ \\
h	-	i\lambda_{k_y}(k_x) +\Delta &\xi_{k_y} (k_x)
\end{array}\right) \non \\
&~& 
\end{eqnarray}
with $\xi_{k_y} (k_x)=  \frac{k_x^2}{2m} -\mu_{k_y}$, with $\mu_{k_y} = \mu-\frac{k_y^2}{2m}$
and $\lambda_{k_y}(k_x)= \lam_I k_x(k_x^2-3k_y^2)  $ and $ h = \frac{1}{2}g\mu_B  B $.
	
The topologically non-trivial phase of $\mathcal{H}_{k_y}^{1D}$ is associated with a
winding of the phase of the determinant of the $\mathcal{Q}$ matrix:
\cite{Zak1989,Ryu2002,Volovik2003,Qi2008,Tewari2012}:
\begin{eqnarray}\label{eq:winding}
\nonumber
W &=& \frac{1}{2\pi }\int_{BZ_{1D}} \partial_{k_x}\mathrm{Im} \log \frac{\det(\mathcal{Q}_{k_y}(k_x))}{|\det(\mathcal{Q}_{k_y}(k_x))|}, \non \\ \non
%	&=& \frac{1}{2\pi  }\int_{BZ_{1D}} \partial_{k_x}\phi_{k_y}(k_x)\\
&=& \frac{1}{2\pi  }\int_{BZ_{1D}} \partial_{k_x}\phi_{k_y}^{(1)}(k_x)+\partial_{k_x}\phi_{k_y}^{(2)}(k_x),\non \\ \non
&=&W_1+W_2,
	\end{eqnarray}
% 	Here 	
% 	\begin{eqnarray}
% 	\nonumber
% 	\tan  \phi_{k_y} (k_x)& =& \frac{2\lambda_{k_y} (k_x) \Delta}{ \xi_{k_y}
%(k_x)^2 -h^2-\lambda_{k_y} (k_x)^2 +\Delta^2},
% 	\end{eqnarray}
where we have used the fact that the winding of the determinant can be expressed as the sum of
the winding of two complex eigenvalues of the $\mathcal{Q}$ matrix:
\begin{eqnarray}
q^{(1,2)}_{k_y}(k_x)=|q^{(1,2)}_{k_y}(k_x)|e^{i\phi_{k_y}^{(1,2)}(k_x)}.
\end{eqnarray}
%hermitian $\mathcal{Q}_{k_y}(k_x)$ matrix as:
%	\begin{eqnarray}
%		W = \frac{1}{2\pi  }\int_{BZ_{1D}} \partial_{k_x}\phi_{k_y}(k_x)
%	\end{eqnarray}
	
\begin{figure}[htb]
\centering
\begin{center}
\sfig[]{\ig[height=5cm]{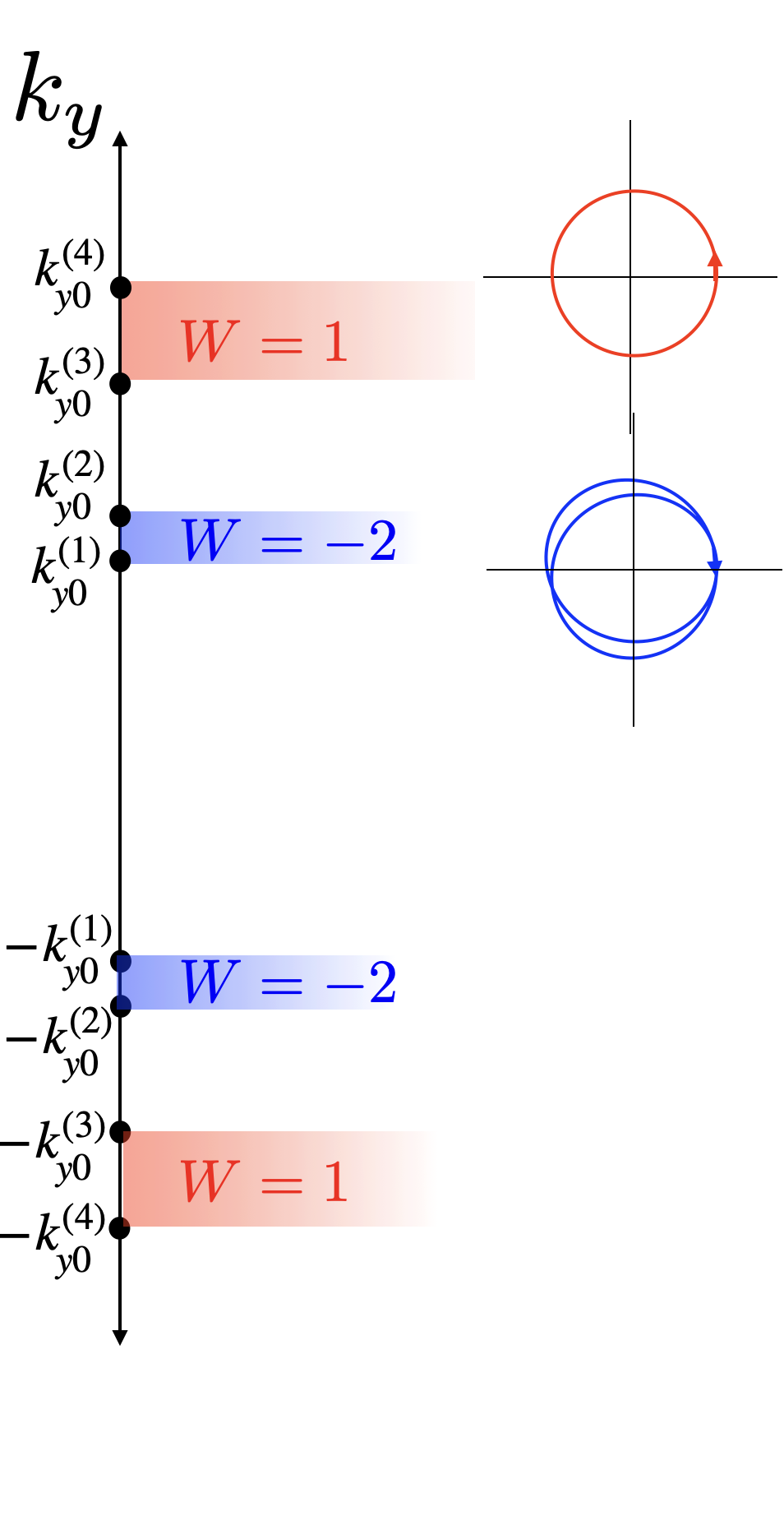}}
\sfig[]{\ig[height=5cm]{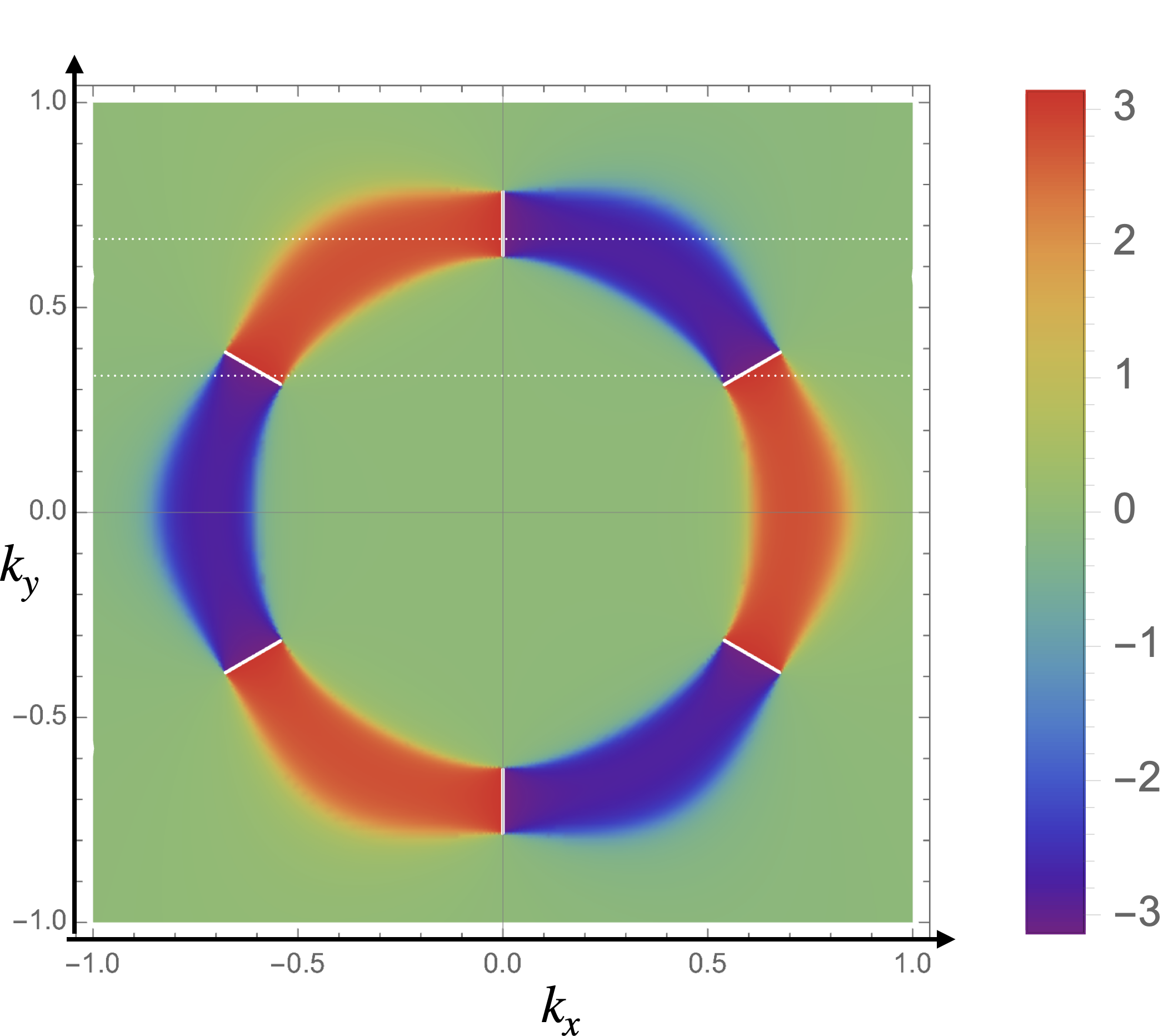}\label{fig:winding_topological}}
\end{center}
\caption{(a) Phase diagram of the family of 1D chiral Hamiltonians as a function of
the transverse momentum parameter $k_y$. The parameter range between the nodal
points $k_{y0}^{(1)}<k_y <k_{y0}^{(2)}$  and  $k_{y0}^{(3)}<k_y <k_{y0}^{(4)}$
corresponds to a topologically non-trivial phase with a topological invariant of
$W =-2 $ and $W=1 $, respectively.  The winding $W$ of the phase of the
determinant of the $\mathcal{Q}$ matrix, Eq.~\eqref{eq:winding}
accumulated as $k_x$ is sweeped over the Brillouin Zone is presented
%MK as a function of the
%longitudinal momentum $k_x $ 
for the different values of $k_y$
%, given by  is depicted 
in Fig (b).}
\label{fig:winding}
\end{figure}

The winding of the phase of the determinant  given by Eq. \eqref{eq:winding} for different
values of $k_y$ is shown in Fig. \ref{fig:winding_topological}. For $k_{y0}^{(1)}<|k_y|<k_{y0}^{(2)}$
(region $\rm{II}$) the phase winds by $-4\pi$ and for $ k_{y0}^{(3)}<|k_y|<k_{y0}^{(4)}$
(region $\rm{IV}$) the phase of the determinant winds by $+2\pi$.
	
\section{Josephson Junction}\label{sec:JJ}
The nodal topological SC phase is characterized by a flat band of Majorana modes
that form on the armchair edges of the sample \cite{He2018}. Here we show that the
non-trivial topology also results in a distinctive current-phase relation when the
system is patterned into a Josephson junction. 
\begin{figure}[htb]
\centering
\sfig{\ig[width=7 cm]{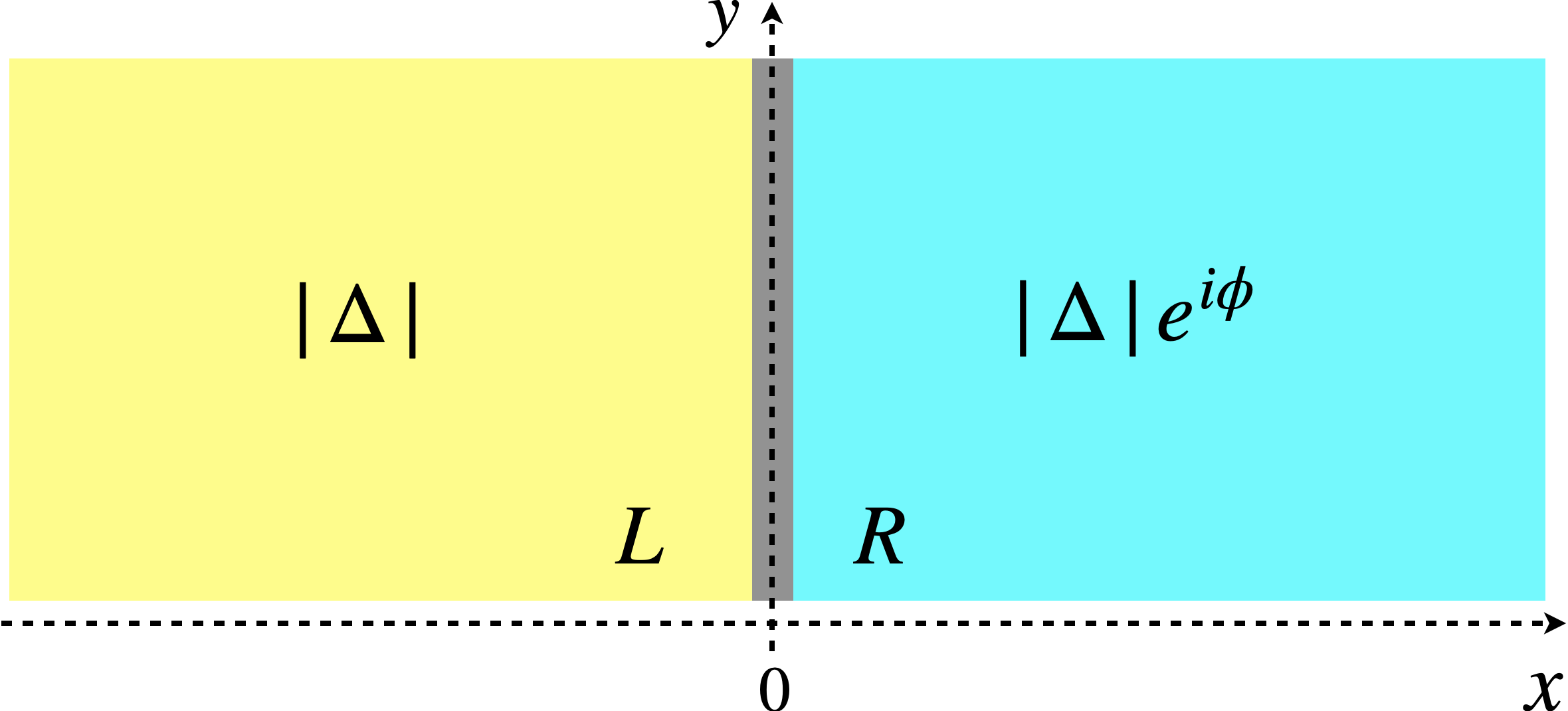}}
%   \sfig{\ig[width=5cm]{Vx.pdf}}
\caption{Schematic showing the Josephson junction between two nodal SCs. The pairing
$\Delta$ is in general complex and differs in phase between the two sides ($L$ and $R$)
of the junction as given by Eq. \eqref{eq:JJSc}. The two sides are separated by a
$\delta-$function barrier given in Eq. \eqref{eq:JJU}.}
\label{fig:JJscheme}
\end{figure}

To study the current-phase relation we consider a junction between two nodal
SCs as shown in Fig. \ref{fig:JJscheme}. The SCs on either side of the junction
are described by a bulk Hamiltonian of the form given in Eq.\eqref{eq:Hkk}.
The superconducting pairing has the same amplitude $|\Delta|$, but differs on
either of the junction by a phase, i.e.
\bes
\beq
\Delta(x) = \begin{cases} |\Delta|, &\text{for $ x < 0$}\\
|\Delta| e^{i\phi}, &\text{for $x > 0$.} 
\end{cases}\label{eq:JJSc}
\eeq
In addition, there is a $\delta-$function barrier at $x=0$
\beq
U(x) = U_0 \delta(x). \label{eq:JJU}
\eeq
Since the translational invariance along the $x-$direction is broken, we replace 
$k_x \rightarrow -i \partial_x$. However, since the translation symmetry along
the $y-$direction is preserved, $k_y$ is still a good quantum number. Therefore,
$k_y$ can be treated as a parameter and the Josephson-junction problem can be
solved independently for each $k_y$. Written in this form, Eq. \eqref{eq:Hkk}
becomes,
\bea
\mathcal{H}_{k_y}(x)&=&\Big(-\frac{1}{2m}{\partial^2_x}- \mu_{k_y}+U(x)\Big)\tau^z \non \\ \non \\
&+& i\lam(\partial^3_x+3 k_y^2 \partial_x) \sig^z+ H \tau^z \sig^x \non \\ \non\\
&+&\Re(\Delta(x)) \tau^y \sig^y +\Im(\Delta(x)) \tau^x \sig^y \label{eq:Hxk}
\eea
\ees
The wave function $\Psi(x)$ for a given value of $k_y$ satisfies the eigenvalue
equation
\beq
\mathcal{H}_{k_y}(x) \Psi(x) = E_{k_y} \Psi(x), \label{eq:eigenmodes}
\eeq
with the Hamiltonian given by Eq. \eqref{eq:JJSc} - \eqref{eq:Hxk}.
Here $\Psi(x) =(u_{\uparrow}(x),u_{\downarrow}(x),v_{\uparrow}(x),v_{\downarrow}(x))^T$
is a four-component spinor, and we have suppressed the dependence of the wave function
on the $k_y$ parameter for brevity. Let the wave function to the left and right of the
barrier (labelled as ``L''  and ``R'' in Fig. \ref{fig:JJscheme}) be denoted by
$\Psi_L(x) $ and $\Psi_R(x) $  respectively. For the wave function to be
continuous and differentiable at the junction, the following boundary conditions
hold at $x=0$
\bes
\bea
\Psi_L(0)  &=& \Psi_R(0)  \label{eq:con}  \\ 
&~& \non\\
\left.\frac{\partial\Psi_R(x)}{\partial x}\right|_{0}-\left.\frac{\partial\Psi_L(x)}{\partial x}\right|_{0}
&=& 2 \alpha \Psi(0)\label{eq:diff}
\eea
where  $\Psi(0) = (\Psi_L(0) + \Psi_R(0))/2 $ and $\alpha = mU_0$. 
Also, we define the transparency $D$ of the barrier as,
\beq
D = \frac{1}{\al^2 + 1}\label{eq:JJD}
\eeq
\ees
such that a transparent junction corresponds to $D=1 $, while an opaque barrier is given by
$D\rightarrow 0 $. 

A particle with a given energy  $E$ and transverse momentum $k_y$ propagates with
linear momentum $k_x(E,k_y)$ which is determined by inverting the dispersion
relation.  In the four-band model we consider, for every pair of $E$ and $k_y$,
there are four possible values of $k_x$ which in the presence of a superconducting
gap are in general complex (i.e. have an oscillatory as well as a decaying component). 

At energies below the induced superconducting gap $E<\Delta_{\rm gap} $, the wave
function decays exponentially away from the junction. Therefore, general solutions of 
Eq. \eqref{eq:eigenmodes} to the left/right of the barrier can be expressed as:
%For the 
%solution to be physical, we need the wave function %to vanish at $\pm \infty$.
%Therefore on the left of the barrier, since we require the wave function
%$\Psi_L$ to decay as $x\rightarrow -\infty$, we need $Im(k_x)<0$;
%Similarly, to the right of the barrier, as the wave function $\Psi_R$ to
%decay as $x\rightarrow +\infty$, we need $Im(k_x)>0$. We can then
%write $\Psi_L$ and $\Psi_R$ as linear combinations of the eigenstates chosen
%according to the above conditions,
\bes
\bea
\Psi_L(x) = \sum_{j=1}^4A_{j} \psi_{k_j} e^{i k_{j} x}, \\
\Psi_R(x) = \sum_{j=1}^4B_{j} \bar{\psi}_{k_j}e^{i\bar{k}_{j} x},
\eea
\label{eq:psilr}
with ${\rm Im}(k_j)<0$ and ${\rm Im}(\bar{k_j})>0$.
Here each wave-function is a four-component spinor, i.e. $\psi_{k_j} = (u_{k_j,\uparrow},u_{k_j,\downarrow},v_{k_j,\uparrow},v_{k_j,\downarrow})^T$.
\ees

Substituting the expression for the wave function in Eq. \eqref{eq:psilr},
the two boundary conditions Eq. \eqref{eq:con} and \eqref{eq:diff} give us the
following matrix equation for the coefficients $A_j $ and $B_j$,
\bes
\beq
M X = 0,  \label{eq:MX0}
\eeq
where X is the column vector formed by $A_j$s and $B_j$s. The matrix $M$ is 
constructed using the four $k_j$ momenta and the corresponding four-spinors 
$\psi_{k_j}$ as follows,
\beq
M = \begin{pmatrix} \psi & & -\bar{\psi} \\  \\ -\psi 
(iK+\al\mathrm{I}_{4\times4}) & 
&\bar{\psi}(i\bar{K}-\al\mathrm{I}_{4\times4}) \end{pmatrix} \label{eq:M}
\eeq
\ees
Here $\psi$ and $\bar{\psi}$ are the four matrices formed by the column vectors 
$\psi_{k_j}$ and $\bar{\psi}_{k_j}$ respectively. $K$ is a $4\times 4 $
diagonal matrix formed by the momenta $k_j$s, $K={\rm diag} (k_1,k_2,k_3,k_4)$.
%$$K = \begin{pmatrix}	k_1 & 0& 0& 0\\		0&	k_2 %& 0& 0\\	0&0&	k_3 & 
%0&\\ 	0 & 0& 0& k_4 
%%\end{pmatrix}$$
%\ees
Solutions to Eq. \eqref{eq:MX0} exist provided the determinant of $M$ vanishes.
From this requirement we obtain the equation for the energies of the Andreev bound
states as a function of the phase difference across the junction, i.e. $E_J(\phi)$.
Using this we calculate the Josephson current as 
\beq
I_J(\phi) = 2e\frac{\partial E_J}{\partial\phi} \label{eq:IJ}
\eeq
The periodicity of the energy $E_J$ and current $I_J$ tells us about the topological
character of the system. While the non-topological regime shows a periodicity of $2\pi$, in the topological
phase both quantities follow a $4\pi$ periodicity. This is explained in detail in the following section.
\section{Results}\label{sec:res}
 We work in a parameter regime where the in-plane field exceeds the
Pauli limit. In particular, we fix $h_x =0.06$ and $|\Delta| =0.02$. The spectrum,
as noted earlier, is gapless and nodal points appear along $k_x = 0,\pm\sqrt{3}k_y$, i.e.
the lines where the SOC vanishes. The dispersion relation for this set
of parameters is shown in  Fig. \ref{fig:surf_H6Del2}.

\begin{figure}[htb]
\centering
\sfig[Josephson Energy]{\ig[width=4.4cm]{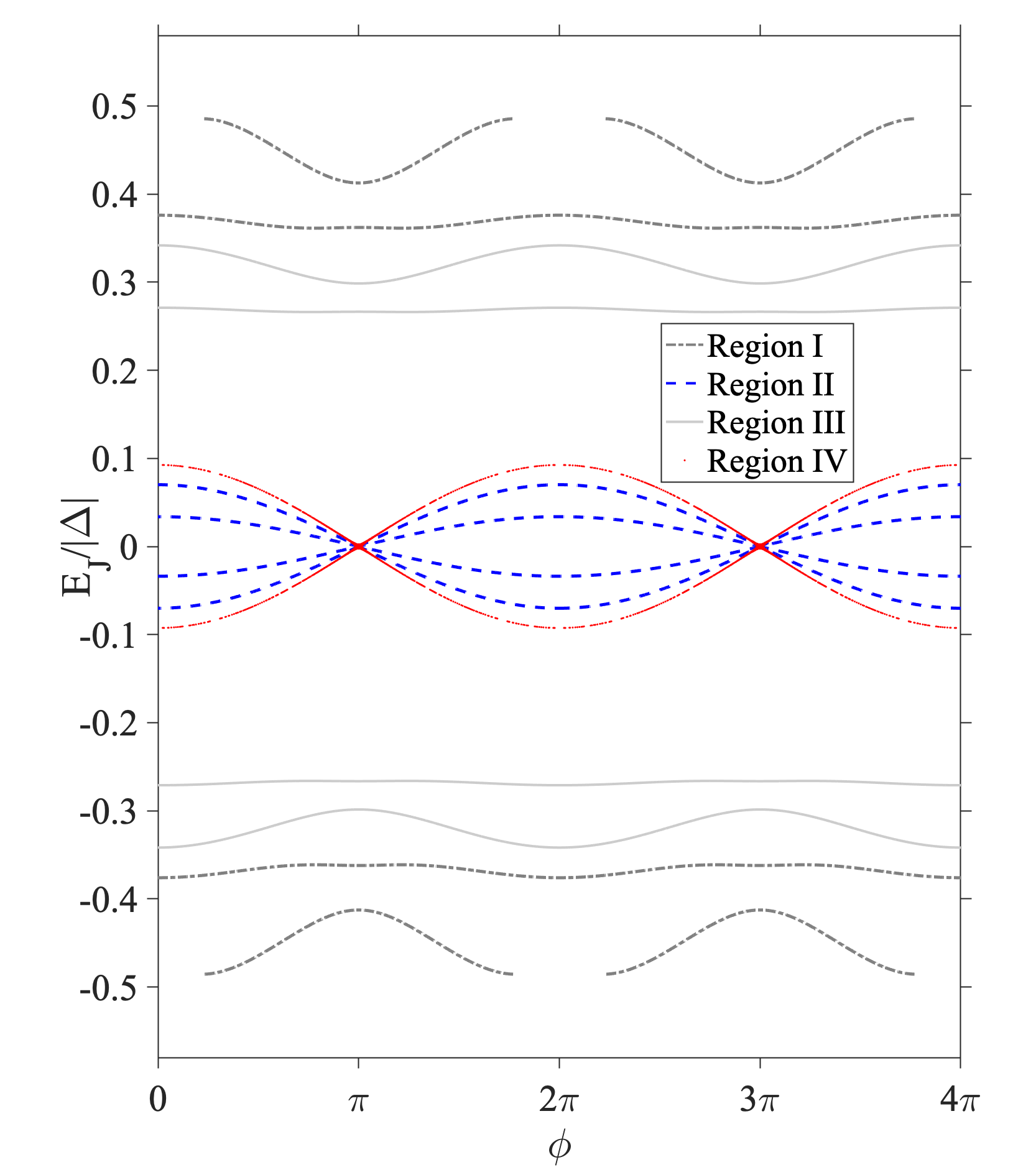}\label{fig:EjD2}}\hspace{-0.4cm}
\sfig[Josephson Current]{\ig[width=4.4cm]{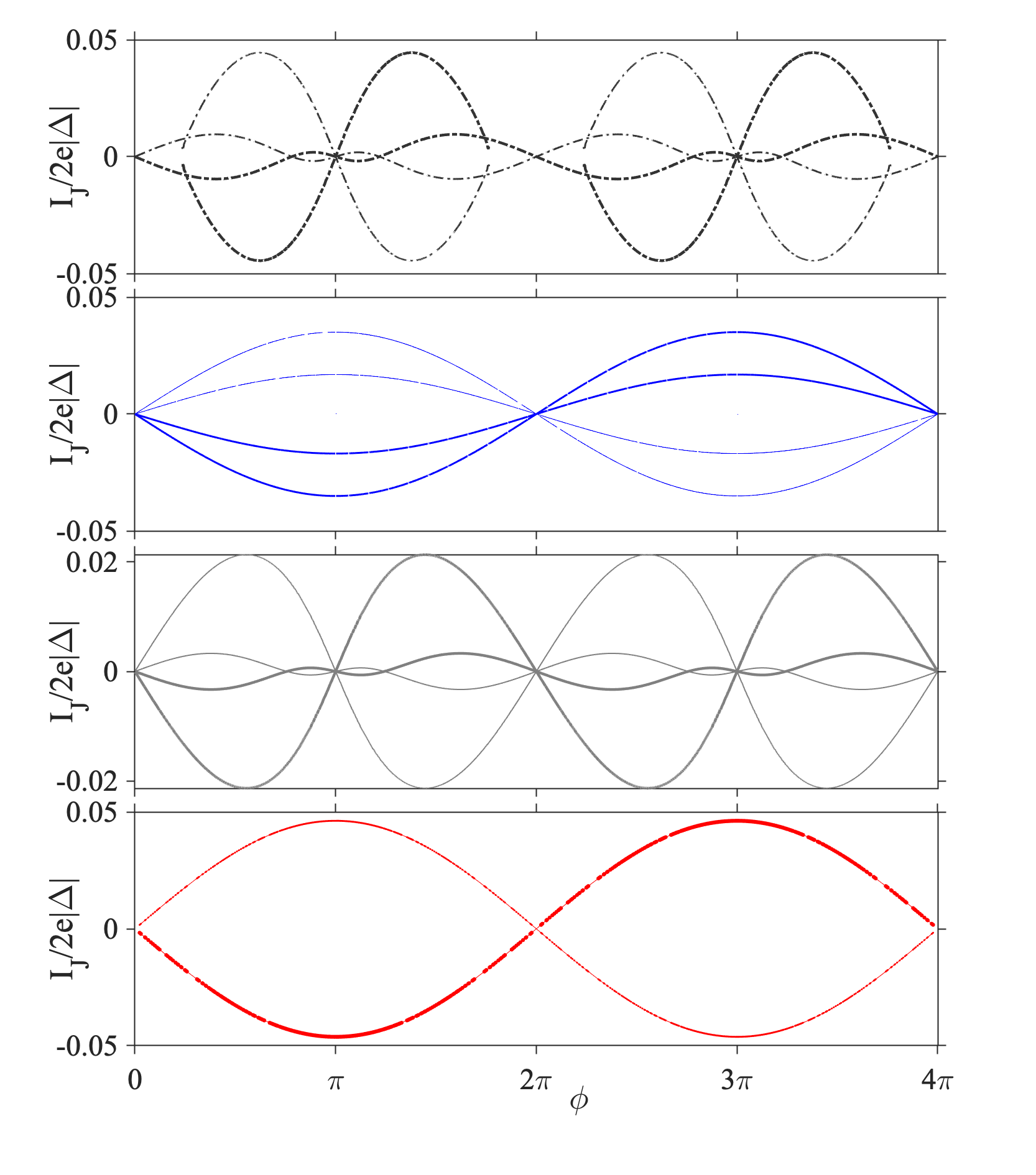}\label{fig:IjD2}}
\caption{(a) Josephson Energy $E_J(\phi)$ and (b) Josephson Current $I_J(\phi)$ for the 
midgap states at representative points in the four different regions labelled $\rm{I - IV}$
in Fig.\ref{fig:surf_H6Del2}. The red (dotted) and blue (dashed) curves lie in the
topological regions $II$ and $IV$ respectively. Clearly, these intersect the $E=0$ line at
$\phi = \pi$ and $3\pi$, and have a periodicity of $4\pi$. On the other hand, the dark gray
(dash-dotted) and light-gray (solid) curves are far away from $E=0$ for all values of $\phi$.
These lie in the non-topological regions $\rm{I}$ and $\rm{III}$ respectively, where the
zero-energy modes are not allowed. Here the periodicity is found to be $2\pi$. These
observations are found to be true for various strengths/transparencies of the barrier
separating the two sides of the Josephson junction.}
\label{fig:D2}
\end{figure}

Figure \ref{fig:EjD2} shows plots of $E_J$ vs $\phi$ along which the determinant of the
matrix $M$ in Eq. \eqref{eq:M} vanishes for barrier transparency $D=0.2$. This means
that it is only along these curves that solutions for Eq. \eqref{eq:MX0} exist. The dark
gray (dash-dot) and light gray (solid) curves belong to the non-topological regions
$\rm{I}$ and $\rm{III}$ respectively. In both these regions we note that there is no
zero-energy mode for any value of the phase $\phi$. Moreover, the periodicity of these
two $E(\phi)$ curves is $2\pi$ as can be seen from the figure. On the other hand, the
red (dotted) and blue (dashed) curves correspond to the two topological regions
$\rm{II}$ and $\rm{IV}$ respectively. As is evident from the figure, these cross
$E=0$ at $\phi=\pi$ and $3\pi$ and have a periodicity of $4\pi$. Even though both these
regions are topological, they differ in terms of the winding number as shown in
Fig. \ref{fig:winding}. The blue (dashed) curve has two branches since the winding number
in the corresponding region $\rm{II}$ ($k_{y0}^{(1)} < k_y < k_{y0}^{(2)}$)
is $-2$; whereas the red (dotted) curve has only one branch as the winding number in region
$\rm{IV}$ ($k_{y0}^{(3)} < k_y < k_{y0}^{(4)}$) is $+1$.

From this, we can also infer the behavior of the Josephson current $I_J$ using
Eq. \eqref{eq:IJ}. This is shown in Fig. \ref{fig:IjD2} as four panels for the four
different regions. In the trivial regions $\rm{I}$ (dark gray, dash-dot)
and $\rm{III}$ (light gray, solid) the current-phase relation has a period of $2\pi$.
The current corresponding to the positive-energy branches is shown as a thick line
whereas the negative-energy branch is shown as a thin line in both cases.
On the other hand, both the topological regions $\rm{II}$ (blue, dash) and $\rm{IV}$
(red, dot) show a $4\pi$ periodicity in the current response. The difference between
these two phases is that while region $\rm{II}$ has two modes (corresponding to winding
number $W=-2$), region $\rm{IV}$ has only one branch ($W=+1$). The thick (thin) curves
in each panel correspond to the branches that have positive (negative) energy at $\phi=0$.

As the transparency $D$ of the barrier is reduced, the mid-gap Andreev states in the
trivial phase of the 1D Hamiltonian flatten and move to energies closer to the
induced gap. On the other hand, for the set of transverse momenta that correspond
to the topological regime, the presence of a pair of zero-energy Majorana modes
at the barrier result in energy levels that stick to zero energy at a phase
difference of $ \phi=\pi$. This dependence on the barrier transparency
is briefly explained in appendix \ref{app:barr}.

\section{Discussion and physical picture}\label{sec:disc}
The behavior of the Josephson current can be understood qualitatively by studying the
nature of the mid-gap states localized at each end of the junction in the weak-coupling
limit. Here, the tunneling Hamiltonian between the two sides of the barrier can be
treated in perturbation theory between otherwise decoupled half planes.  Moreover,
following the discussion in Sec. \ref{sec:top} each decoupled half plane can be
treated as a family of semi infinite 1D wires governed by the Hamiltonian
$\mathcal{H}_{k_y}^{1D}(k_x) $.

\begin{figure}[htb]
\centering
\ig[width=.45\textwidth]{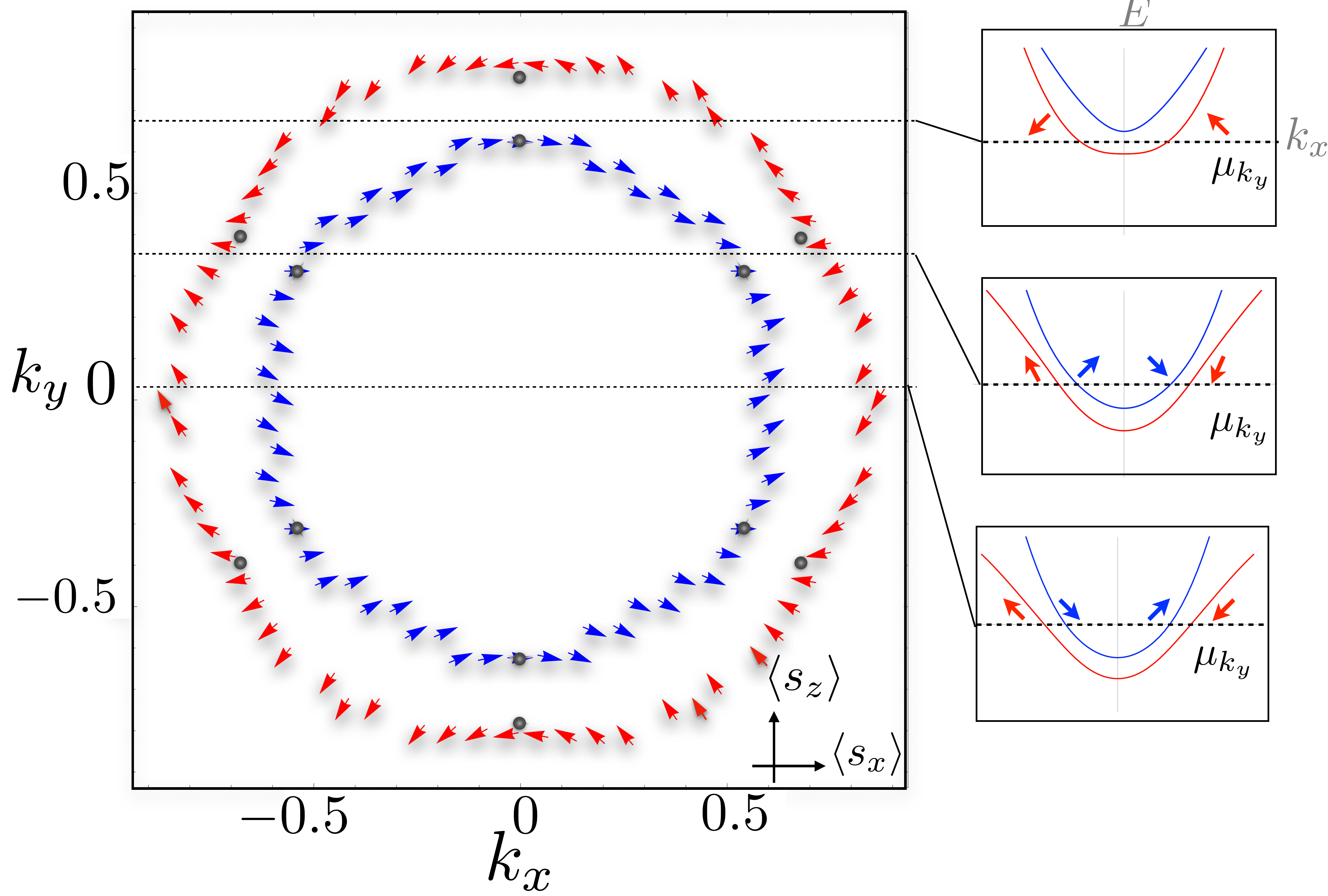}
\caption{The spin texture of the two Fermi surfaces as a function of $k_x $ and $k_y $.
Line cuts at fixed $k_y $ show the dispersion of the two spin channels of the effective
1D model, corresponding to Region \RNum{1}, lower panel, Region \RNum{2}, middle panel
and Region \RNum{4}, upper panel.  }
\label{fig:SpinStructure}
\end{figure}

The  underlying physical picture is depicted in Fig. \ref{fig:SpinStructure} which shows
the spin texture in the normal state $ (\Delta = 0)$ of the two Fermi-surfaces split by
in-plane field and SOC. Different cuts of fixed $k_y $ correspond to
one/two populated bands with normal/flipped spin orientation.  

In the nodal superconducting phase, the condition of $h\gg |\Delta| $ strongly suppresses
inter-spin-channel pairing, and the induced superconductivity is predominantly intra-channel
type. In region \RNum{1} and \RNum{3} corresponding to $|k_y|<k_{y0}^{(1)}$ and $k_{y0}^{(2)}<|k_y|<k_{y0}^{(3)}$ respectively, the  chemical potential of
the one-dimensional wire exceeds the Zeeman splitting $\mu_{k_y}=\mu - k_y^2/2m \gg h_x$ and
the two spin channels acquire intra-band superconducting correlations with opposite winding
numbers $W_{1,2}=\pm1 $, corresponding to a topological trivial phase. The two spin channels
which exist as independent p-wave SCs in the bulk,  are then coupled by the end
of the wire. This coupling gaps out their midgap Majorana-like excitations resulting in a 
finite energy Andreev bound state on either side of the junction. Tunneling across the junction
couples the Andreev states resulting in two modes with a $2\pi$-periodic energy-phase relation. 
   
%In the nodal superconducting phase, the condition of $h\gg |\Delta| $ strongly suppresses
%singlet s-wave
%pairing, and the induced superconductivity is predominantly of p-wave type. In the parameter regime marked
%by the cuts at $k_y=0,k_y^2 $ in Fig. \ref{fig:4bands}, the  chemical potential of the 1D wire exceed the
%Zeeman splitting $\mu_{k_y}=\mu - k_y^2/2m \gg h_x$ and the two spin eigen modes acquire superconducting
%correlations with opposite winding numbers $\nu_{1,2}=\pm1 $. The two decoupled modes are coupled by
% boundary of the wire, which in tern gaps out the midgap Majorana-like excitations at the wire's end
%resulting in a  finite energy Andreev bound state at each side of the junction. The tunneling across the 
%junction couples the Andreev states resulting in two eigenmodes with a $2\pi  $ periodic energy phase
%relation. 

While the qualitative picture of two p-wave channels remains true for region \RNum {2}
($k_{y0}^{(1)}<|k_y|<k_{y0}^{(2)}$), in this parameter regime the spin orientation of the inner
Fermi-surface is flipped with respect to region \RNum{1}, while the outer Fermi-surface remains
unchanged. As a result the two p-wave spin channels have the same winding number $W_{1,2}=-1$.
The two Majorana modes are protected by chiral symmetry and cannot be gapped by boundary.
Consequently, in region \RNum {2} each half wire hosts  two Majorana zero modes at its end.
Treating the tunneling across the junction in perturbation theory would result in two mid-gap
states with an energy-phase relation which is $ 4\pi$ periodic.

Finally, in region \RNum{4}  ($k_{y0}^{(3)}<|k_y|<k_{y0}^{(4)}$), the Zeeman splitting exceed the
critical value $h>\sqrt{\Delta^2+\mu^2} $ and the 1D wire is in the same topological class as
that of the spin orbit nanowire \cite{Oreg2010,Lutchyn2010}. Here only one of the two
spin-channels is populated, leading to a single Majorana zero mode at either
end of the junction.
The tunneling across the junction would then lead to a single midgap state with $4\pi$
periodicity. The different scenarios are discussed in detail in App. \ref{sec:edgemodes}.

%We assume that the electronic momentum parallel to the junction $k_y $ is conserved.
%The two dimensional strip then separates into a set of one dimensional channels 
%with different $k_y$. The total current is expected to be the sum of the different
%channel contributions. Here in addition to the  barrier transparency which depends
%on the $x$ component of the Fermi momentum and will therefore  become channels
%dependent $D(k_y) $, certain channels with $k_y $ values between the nodal points
%would feature a $ 4\pi $ periodic Josephson current, while others will exhibit a
%$2\pi  $ periodicity. Nonetheless we expect that as the barrier becomes more opaque,
%the contribution from topologically trivial channels with $k_y $ away from the
%nodal points would be substantially reduced. 

\section{Conclusion}\label{sec:conclusion}
 We study the current-phase relation of a nodal topological SC in a
Josephson-junction geometry. Despite the presence of an in-plane field, 
% While the presence of an  in-plane field that stabilizes the nodal phase explicitly
% breaks time reversal symmetry $\Theta $, 
the model retains an effective chiral symmetry which arises due to the particle-hole
symmetry and a modified TR symmetry $T= M_z\Theta\tau_Z $. We find that the
Josephson current-phase relation depends on the momentum transverse to the current
direction and shows a distinctive $4\pi $ periodicity when the transverse momentum,
treated as a control parameter, lies in-between pairs of nodal points (regions $\rm{II}$
and $\rm{IV}$). The nodal momenta thus define the boundaries between a trivial phase and two topologically
distinct non-trivial phases characterized by different winding numbers $W=1 $ and $ W=-2$.

The $ W=-2$ phase is protected by the chiral symmetry and is therefore unstable
with respect to symmetry-breaking perturbations such as a Rashba SOC, which couple the
two Majorana flat bands at the boundary. When such perturbations are
present, the Josephson current-phase relation will exhibit a $2\pi $ periodicity 
in region $\RNum{2} $. Conversely, the $W=1 $ phase in region $\RNum 4 $ is stable to
weak perturbations that break the modified TR symmetry. We therefore expect
that the $4\pi$ periodicity associated with this region will persist in the presence
of weak symmetry-breaking perturbations.

\section{Acknowledgements}
The authors would like to thank  Beena Kalisky, Amit Keren and Binghai Yan for fruitful discussions. 
D.M.  acknowledges support from the Israel Science Foundation (ISF) (grant No. 1884/18). 
M.K. and D.M. acknowledge support from the ISF, (grant No. 1251/19).

\appendix

\section{In-plane field and phase transition}\label{app:phase}
\begin{figure}[htb]
\centering
\begin{center}
{\ig[width=7cm]{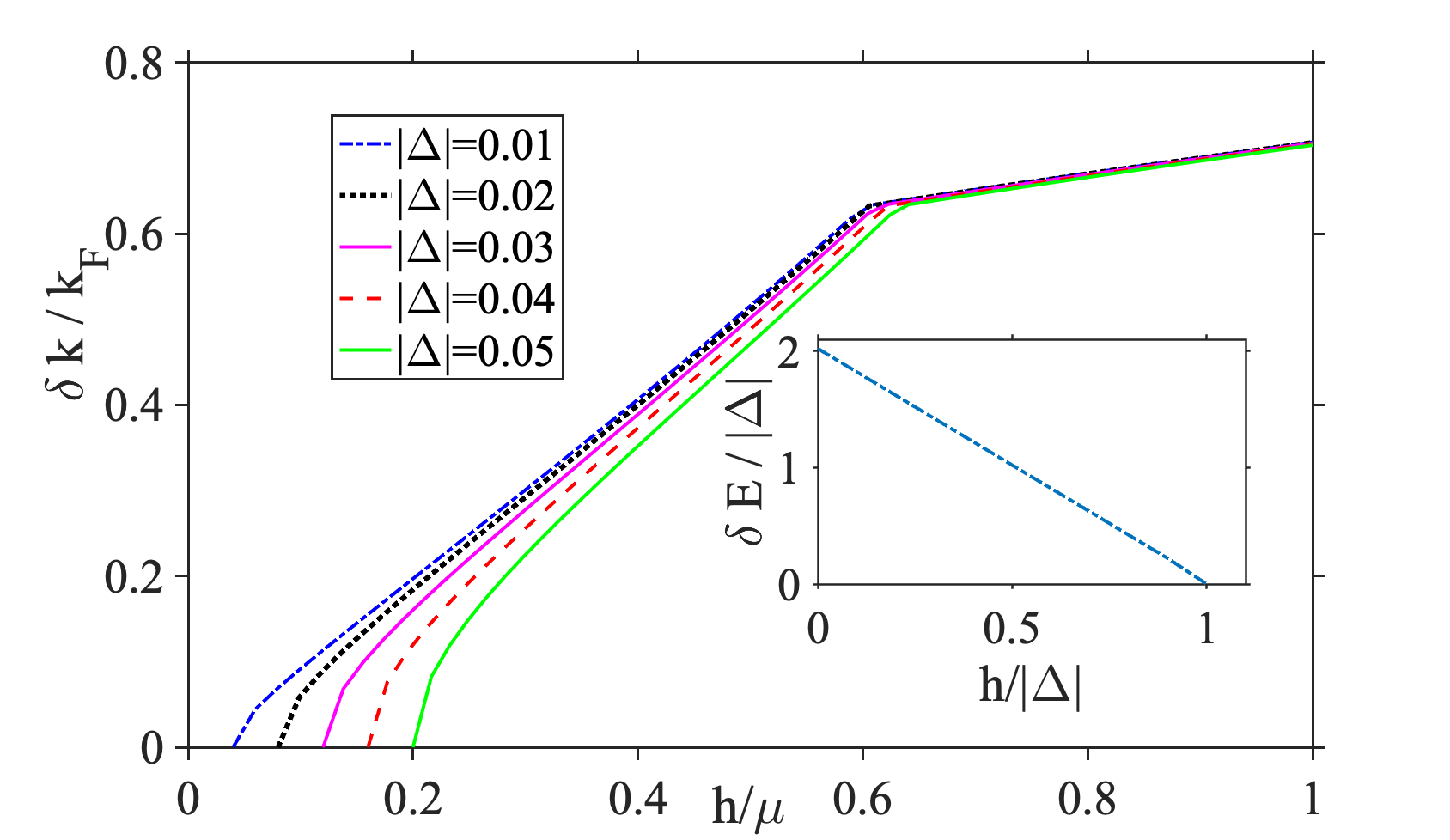}}
\end{center}
\caption{Magnetic field-driven phase transition. The separation between each pair of
nodal points increases as we increase the in-plane field beyond the superconducting
pairing $\Delta$. This is shown for a range of $\Delta$s. The inset shows that the
spectral gap $\delta E$ decreases linearly with increasing $h$. This is true for all
$\Delta$s. The spectrum becomes gapless at $h=|\Delta|$ and continues to remain so
for $h>|\Delta|$.}
\label{fig:separation_h}
\end{figure}

A remarkable property of nodal SCs is that the superconducting properties
survive even when $h$, the applied in-plane magnetic field, is increased beyond the
Pauli limit. At magnetic fields lower than the superconducting pairing, i.e. when
$h<|\Delta|$ the spectrum is gapped at all momenta. As we increase the magnetic field,
this energy gap reduces linearly as shown in Fig.\ref{fig:separation_h} (inset). This
trend holds true for all $\Delta$s.

At $h=|\Delta|$ the system undergoes a phase transition which is accompanied by a
closing of the spectral gap at six points in the Brillouin zone (B.Z.). These six
gapless points or `nodes' lie at the vertices of a regular hexagon. These mark the
points where the Fermi surface $|k_f|= \sqrt{2 m \mu}$ intersects the $\lam(\vk)=0$
lines. 

As $h$ is increased beyond the pairing $|\Delta|$ the spectrum continues to remain
gapless. However, with increasing $h$, each of the six nodal points splits into two.
This means that each B.Z. hosts a total of twelve nodal points which are shown in
Fig.\ref{fig:surf_H6Del2}. Each of the six pairs of nodal points continues to spread
as $h$ is increased further. The separation of the nodes as a function of $h$ is shown
in Fig.\ref{fig:separation_h} for a range of $\Delta$s.

% As explained in Sec. \ref{sec:BdG} the low energy Hamiltonian for a nodal
%superconductor is given by
% $$\mathcal{H} = \xi(\vk) \tau^z   \sig^0 + \lambda(\vk) \tau^0 \sig^z+ h \tau^z  
% \sig^x +\Delta \tau^y   \sig^y.$$
% The Ising spin-orbit coupling, $\lambda(\vk) = \lam_I(k_x^3-3 k_x k_y^2)$
% vanishes along the lines  $k_x=0, \pm \sqrt{3}k_y$.
% The energy spectra are shown in Fig.\ref{fig:4bandsscan}
% When $h<|\Delta|$, the spectrum is gapped as shown in 

%\input{appendix_2}
\section{Edge states of the effective 1D Hamiltonian}\label{sec:edgemodes}

We study the nature of the mid-gap states localized at each end of the junction in the
weak-coupling limit, where, the tunneling Hamiltonian between the two sides of the barrier
can be treated in perturbation theory between otherwise decoupled half wires. The underlying
physical picture is depicted in Fig. \ref{fig:SpinStructure} which shows the spin texture in
the normal state ($ \Delta = 0$) of the two Fermi surfaces split by in-plane field and
SOC. Different cuts of fixed $k_y $ correspond to one/two populated
spin-channels  with normal/flipped spin direction. Different scenarios are discussed in
detail below. 

\subsection{Region \RNum {1}: $|k_y|<k_{y0}^{(1)}$}
This regime of parameters is characterized by two occupied spin-channels (see lower panel
in Fig. \ref{fig:SpinStructure}). The condition, $\mu= k_F^2/2m \gg h,|\Delta| $ allows us
to linearize the spectrum near the Fermi points. The general form of the wave function
is given by,
\begin{equation}
\psi_{\sigma=\uparrow,\downarrow} = R_\sigma(x)e^{ik_F x}+L_\sigma(x)e^{-ik_F x}
\end{equation}
where $R_\sigma(x)/L_\sigma(x) $ are slowly varying functions of $x$ describing right/left
moving electrons. In the basis of the slow varying fields $\psi_{+} = (R_\uparrow(x),
R_\downarrow(x),L_\uparrow(x)^\dag,L_\downarrow(x)^\dag $) and $\psi_{-} = (L_\uparrow(x),
L_\downarrow(x),R_\uparrow(x)^\dag,R_\downarrow(x)^\dag) $, the BdG hamiltonian is:
\begin{equation}
H_{k_y,\pm}^{1D}(x) = \mp iv_F\partial_x \tau_z+\pm\lambda k_F^3\sigma_z + h \sigma_x\tau_z +
\Delta \sigma_y\tau_y.
\end{equation}
We perform a rotation in the spin space to diagonalize the particle-conserving terms
by applying the unitary transformation, $\tilde{H}_{k_y,\pm}^{1D}(x)=U^{\dag} H_{k_y,\pm}^{1D}(x) U$,
where $U = \exp \left[- i \sigma_y \tau_z (\pi/4 - \theta/2) \right]$ with the rotation angle $\theta$
defined via 
\bea
\nonumber
v_Fk_m\cos\theta&=&h\\
v_Fk_m \sin\theta &=& \lambda k_F^3.
\eea
We find $\tilde{H}_{k_y,\pm}^{1D}(x)=H_\pm^{0}+\delta H$ where
\bea
\nonumber
H_\pm^{0} &=& \mp iv_F\partial_x \tau_z+v_Fk_m\sigma_z \pm\Delta \sin \theta \sigma_y\tau_y \\
\delta H&=& \Delta \cos \theta \tau_x
\eea
and we have dropped the subscript $k_y $ for brevity. 
In the limit $h\gg \Delta $ we can treat $\delta H $ in perturbation theory. 
The unperturbed Hamiltonian $H_\pm^0 $ admits two zero energy solutions at the end of the wire:
\bea
\nonumber
\psi_{\pm\downarrow}(x) &=& \frac{1}{\Omega}\left(\begin{array}{c}
     \beta   e^{i\frac{\pi}{4}+i\frac{\phi(x)}{2}}  \\
     0\\
       0\\
      e^{-i\frac{\pi}{4}-i\frac{\phi(x)}{2}}
    \end{array}\right) e^{\pm i (k_F-k_m) x}e^{\beta x/\xi }\\
 \psi_{\pm\uparrow}(x) &=& \frac{1}{\Omega}\left(\begin{array}{c}
    0 \\
         \beta  e^{-i\frac{\pi}{4}+i\frac{\phi(x)}{2}} \\
       e^{i\frac{\pi}{4}-i\frac{\phi(x)}{2}}\\
     0
    \end{array}\right) e^{\pm i (k_F+k_m) x}e^{\beta x/\xi }~~~
\eea
where $\xi^{-1}= \Delta\sin\theta/v_F  $ and $\beta =\pm $ for the left/right side of the junction,
respectively, and 
\bea
\phi(x) = \left\{ \begin{array}{c c}
    0 & x<0  \\
    \phi &x>0 .
\end{array}\right.
\eea
From these zero-energy solutions we can construct two zero-energy modes on each side of the
barrier that satisfy the boundary conditions $\psi_M(x=0)=0 $ namely:
\bea
\nonumber
\phi_{\downarrow\beta}(x) &=&\frac{1}{\Omega} \left(\begin{array}{c}
     \beta   e^{i\frac{\pi}{4}+i\frac{\phi(x)}{2}}  \\
     0\\
       0\\
      e^{-i\frac{\pi}{4}-i\frac{\phi(x)}{2}}
    \end{array}\right) \sin[(k_F-k_m)x]e^{\beta x/\xi }\\
 \phi_{\uparrow\beta}(x) &=& \frac{1}{\Omega}\left(\begin{array}{c}
      0 \\
         \beta  e^{-i\frac{\pi}{4}+i\frac{\phi(x)}{2}} \\
       e^{i\frac{\pi}{4}-i\frac{\phi(x)}{2}}\\
     0
    \end{array}\right) \sin[(k_F+k_m)x]e^{\beta x/\xi}~~~~~ 
\eea

Calculating the matrix element due to the local inter-band pairing term, $\delta H$ between
these states, we find that the  each semi-infinite wire on either side of the barrier hosts
a single Andreev end state with energy: 
\bea
\nonumber
\beta\Delta_{gap} &=& \int dx \langle \phi_{M\uparrow\beta}(x)| \delta H | \phi_{\downarrow\beta}(x)\rangle\\
&\approx&2\beta \frac{\Delta^3}{k_m^2} \cos\theta \sin^2\theta
\eea
Where we have used $k_F\gg k_m\gg\Delta $.
%The tunneling matrix elements can be found following \onlinecite{Kwon2004} and are given by:
%\bea
%\nonumber
% E_{J\sigma} &=&\int dx \langle \psi_{M \sigma L}(x)|  H_t | \psi_{M\sigma R}(x)\rangle\\
% &=&\Delta \sin\theta \cos \phi/2 \sqrt{D_\sigma}
%\eea
%where $\sqrt{D_\sigma} = 2\sqrt{\tau} \sin{(k_F+\sigma k_m)l}/v_F $ is the spin dependent 
%transmission amplitude, $\tau  $ is the tunnel coupling, and $l $ is the junction spacing. 

%Diagonalizing the two perturbative coupling in the low energy basis of the four Majorana
%modes we find the energy of the two Andreev bound states:
%\bea
%\nonumber
%E_{\pm}(\phi)= \frac{E_{J\downarrow}-E_{J\uparrow}}{2}\pm \sqrt{\left(\frac{E_{J\uparrow}+E_{J\downarrow}}{2}\right)^2+\Delta_{gap}^2}
%\eea

\subsection{Region \RNum{2}, $k_{y0}^{(1)}<|k_y|<k_{y0}^{(2)}$}
Once more the condition  $\mu_y=\mu - k_y^2/2m= k_F^2/2m \gg h,\Delta $ is satisfied, corresponding
to two filled spin bands. The situation in this regime is similar to that in region \RNum{1} with the
spin orientation of the inner spin channel  flipped while the outer Fermi surface remains unchanged,
see Fig. \ref{fig:SpinStructure} middle panel. The reason for the flip is  that for the parameter
range $k_y $ between the two nodal points  $k_y= \frac{1}{2}(k_{y0}^1+k_{y0}^2)
\approx\sqrt{\frac{m \mu}{2}}$, the spin orbit term vanishes:
\begin{equation}
\lambda_I k_F(k_F^2 -3k_y^2)= 0
\end{equation}
The magnetic field splits the two Fermi points by an amount $ \sim \pm k_m$. As a result of this
splitting, the two spin channels experience a finite SOC of opposite strength.

In the basis of the slow varying fields 
$\psi_{+}= (R_\uparrow(x),R_\downarrow(x),L_\uparrow(x)^\dag,L_\downarrow(x)^\dag $) and $\psi_{-} = (L_\uparrow(x),L_\downarrow(x),R_\uparrow(x)^\dag,R_\downarrow(x)^\dag) $ the BdG Hamiltonan is:
\begin{equation}
H_{k_y\pm}^{1D}(x) =\mp iv_F\partial_x \tau_z-i2 m^2\lambda v_F
\partial_x\sigma_z +h \sigma_x\tau_z + \Delta \sigma_y\tau_y 
\end{equation}
Next we will assume the eigenvectors have an oscillatory part and a slowly varying part: 
\begin{equation}
    \psi_{\pm s}(x) = e^{ \pm s i k_m x}e^{\beta x/\xi}
    \left(\begin{array}{c}
       R_\uparrow   \\
       R_\downarrow\\
       L_\uparrow^\dag\\
         L_\downarrow^\dag
    \end{array}\right)
\end{equation}
%\begin{equation}
%    \psi_{\pm,\downarrow}(x) = e^{\pm i k_m x}e^{\beta x/\xi}
%    \left(\begin{array}{c}
%       R_\uparrow   \\
%       R_\downarrow\\
 %      L_\uparrow^\dag\\
 %        L_\downarrow^\dag
 %   \end{array}\right)
%\end{equation}
where $s=\pm $ correspond to the two spin eigenvalues.

The Hamiltonian in the 4-spinor basis becomes:
\bea
\nonumber
H_{k_y\pm,s}^{1D} &=& ( s v_F k_m \mp i\beta v_F/\xi)\tau_z \sigma_0\pm 2 m^2\lambda v_F s k_m\tau_0\sigma_z\\
&+& h \sigma_x\tau_z + \Delta \sigma_y\tau_y 
\eea
performing a rotation in the spin basis  $U = \exp \left[- i \sigma_y \tau_z (\pi/4 - s \theta/2) \right]$
this gives rise to $\tilde{H}_{k_y\pm,s}^{1D}  = H_{\pm s}^{0}+\delta H^{1}$:

\bea
\nonumber
H_{\pm s}^{0} &=&  ( s v_F k_m \mp i\beta v_F/\xi)  \tau_z + v_F k_m  \sigma_z \pm s\Delta \sin\theta\sigma_y\tau_y \\
\delta H_{s}^{1}&=&\Delta \cos\theta\tau_x
\eea
where  $v_F k_m \cos\theta =h$ and $v_F k_m\sin\theta = m^2\lambda v_F k_m $.
We therefore identify two zero energy solutions of the form
\bea
\nonumber
\phi_{\uparrow\beta} &=& \frac{1}{\Omega}\left(\begin{array}{c}
      0   \\
       \beta e^{-i\frac{\pi}{4}+i\frac{\phi(x)}{2}} \\
       e^{i\frac{\pi}{4}-i\frac{\phi(x)}{2}}\\
      0
    \end{array}\right)\sin[(k_F+k_m)x]e^{\beta x/\xi } \\
 \phi_{\downarrow\beta} &=& \frac{1}{\Omega}\left(\begin{array}{c}
      \beta e^{-i\frac{\pi}{4}+i\frac{\phi(x)}{2}} \\
       0\\
       0\\
      e^{i\frac{\pi}{4}-i\frac{\phi(x)}{2} }
    \end{array}\right)\sin[(k_F-k_m)x]e^{\beta x/\xi } ~~~~~.
\eea
crucially, the inter band pairing term does not couple the two zero mode: 
\bea
\langle \phi_{\uparrow\beta} | \delta H |\phi_{\downarrow\beta} \rangle=0.
\eea
Consequently, for this regime of parameters, each wire's end hosts two decoupled
Majorana zero modes. Treating the tunneling Hamiltonian in perturbation theory
\cite{Kwon2004} would give rise to two Andreev bound states:
\bea
E_{Js} = \Delta \sqrt{D_s}\sin \theta \cos\phi/2. 
\eea
\\

We note that the presence of the two Majorana modes is protected by the modified TR symmetry
$T$, and is unstable to symmetry-breaking perturbation such as Rashba SOC, the presence of which
would couple the two Majorana modes resulting in a single Andreev state on either side of the barrier.
Consequently, when  this $T$ symmetry is broken, this phase is continuously connected to region
\RNum{1} and the nodal points along the $|k_x|=\sqrt{3}k_y $ line that separate the two will be
gapped out.

\subsection{Region \RNum {4}, $k_{y0}^{(3)}<|k_y|<k_{y0}^{(4)}$}
In this regime of parameters $h> \sqrt{\Delta^2+\mu_{k_y}^2} $ and the effective
1D Hamiltonian is in the topological class of the case of the spin-orbit coupled
nanowire \cite{Oreg2010, Lutchyn2010}. Here $\mu_{k_y}=\mu-k_y^2/2m $.  In region
\RNum {4} only one of the two spin orbit bands is populated and each semi-infinite
half wire hosts a single Majorana zero mode at its end.
Treating the tunneling Hamiltonian in perturbation theory following ref. 
\onlinecite{Kwon2004} would give rise to a single Andreev bound states:
\bea
E_{J} = \Delta \sqrt{D}\sin \theta \cos\phi/2.
\eea

Unlike region \RNum{2}, the presence of a single Majorana mode at either end of the
junction remains stable to a weak symmetry-breaking perturbation.

\section{Dependence on barrier transparency $D$}\label{app:barr}
The behaviour of Josephson energy $E_J(\phi)$ and current $I_J(\phi)$ depend not only on the $k_y$ cut
we choose, but also on the transparency of the barrier that separates the two sides of the junction
shown in Fig. \ref{fig:JJscheme}. We see that for a transparent barrier i.e. when $D\rightarrow 1$ the
Josephson energies are close to zero at $\phi=\pi,3\pi$ in all the regions. However, as we increase
the strength of the barrier, i.e. as $D$ is reduced, the modes in the non-topological regions i.e.
$\rm{I}$ and $\rm{III}$ are gapped out and move away from zero. However the modes in the topological
regions $\rm{II}$ and $\rm{IV}$ continue to exist close to zero energy at $\phi=\pi,3\pi$.
	\begin{figure}[htb]
		\centering
		\sfig[Josephson Energy]{\ig[width=4cm]{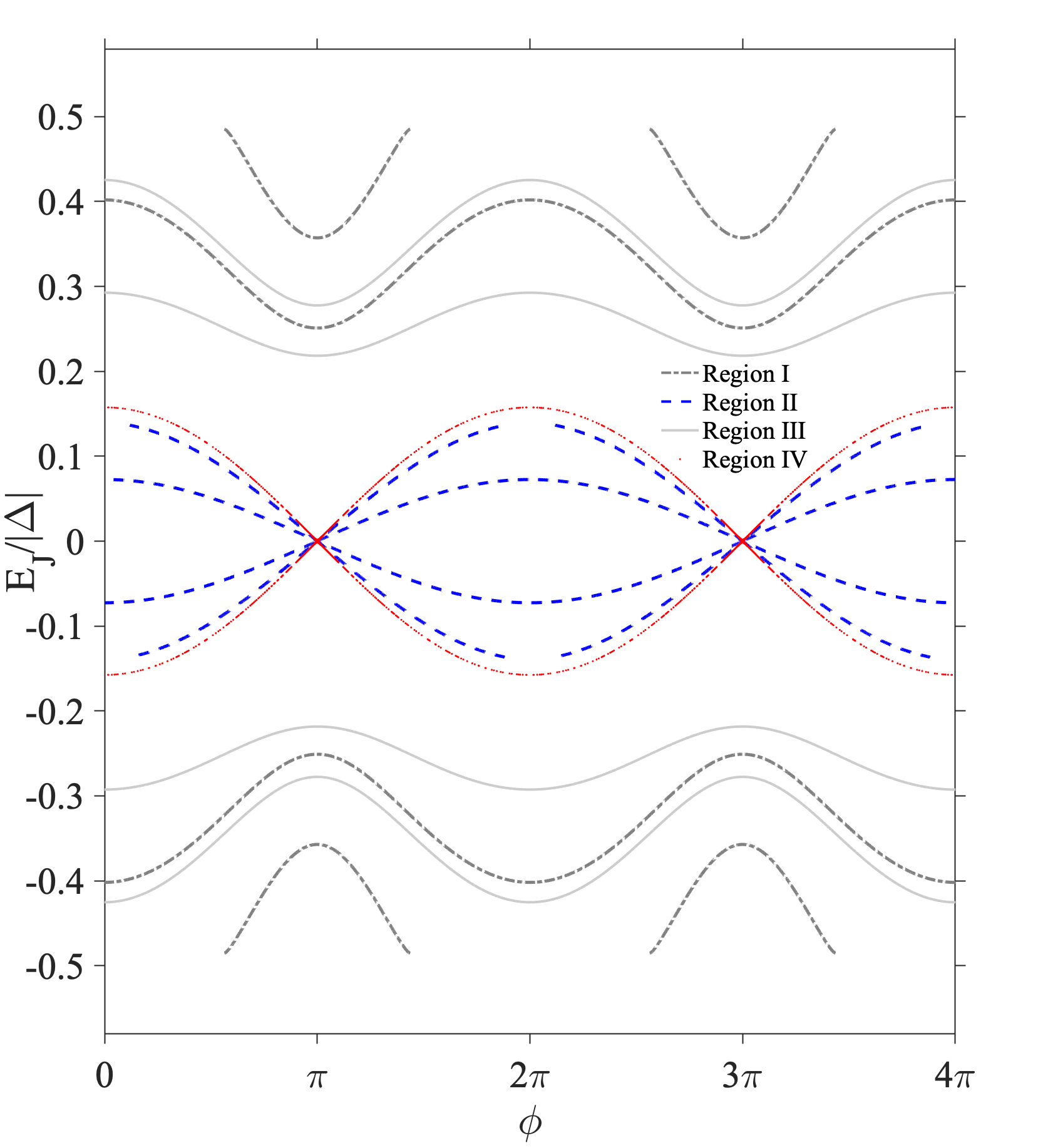}\label{fig:EjD6}}
		\sfig[Josephson Current]{\ig[width=4cm]{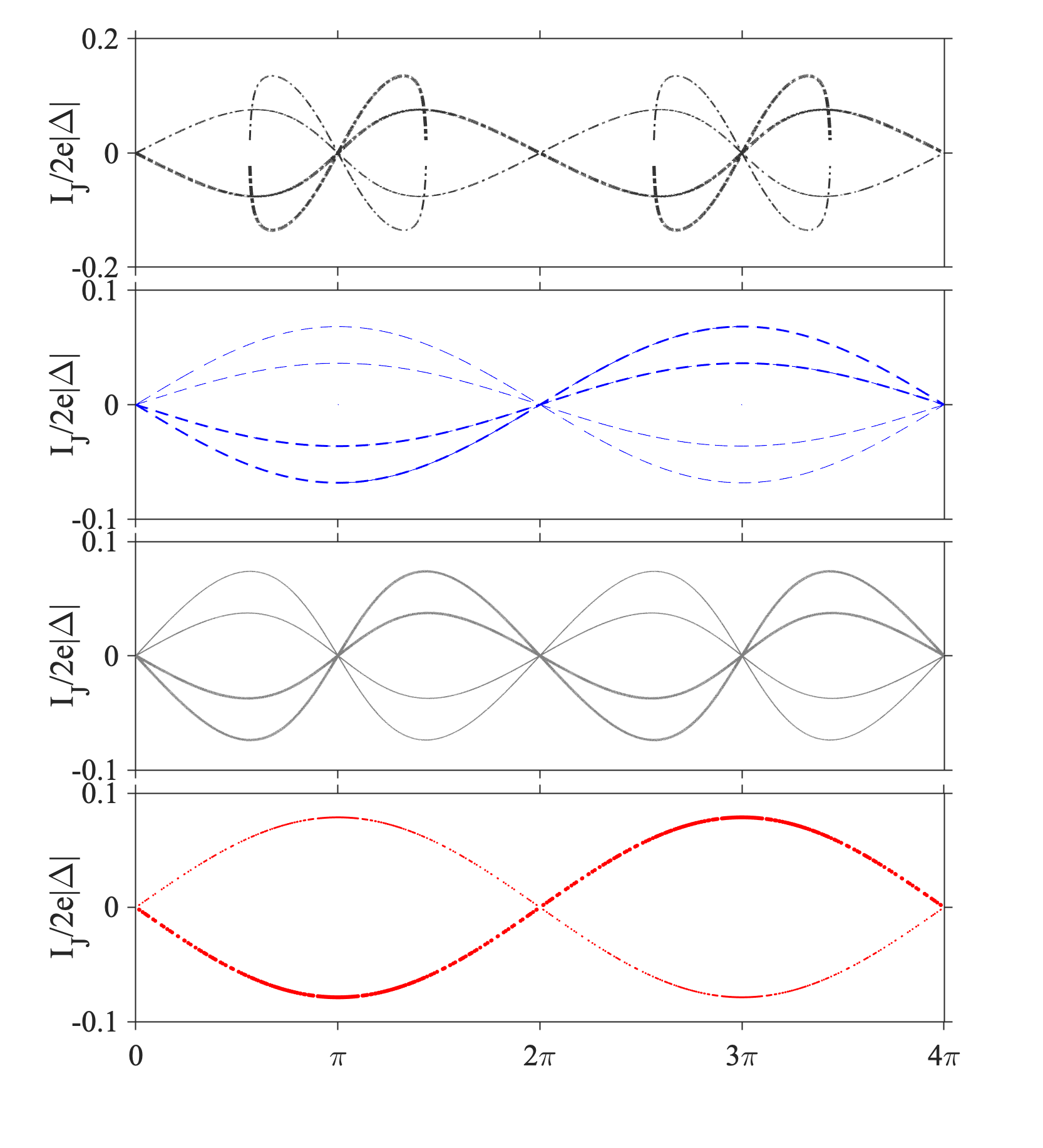}\label{fig:IjD6}}
		\caption{(a) Josephson energy $E_J(\phi)$ and (b) Josephson current $I_J(\phi)$
		with barrier transparency $D=0.6$. The colors correspond to the transverse
		momentum $k_y$ lying in the regions $\rm{I-IV}$.}
		\label{fig:D6}
	\end{figure}
\bibliography{refs}
\end{document}